\documentclass[twocolumn,showpacs,preprintnumbers,amsmath,amssymb]{revtex4}

\usepackage{graphicx}
\usepackage{dcolumn}
\usepackage{bm}

\begin{document}


\title{The Physical Origins of Entropy Production, Free Energy Dissipation and Their Mathematical Representations}

\author{Hao Ge$^1$}
\email{gehao@fudan.edu.cn}
\author{Hong Qian$^{2,1,}$}%
\email{qian@amath.washington.edu} \affiliation{$^1$School of
Mathematical Sciences and Centre for Computational Systems Biology,
Fudan University, Shanghai 200433, PRC. $^2$Department of Applied
Mathematics, University of Washington, Seattle, WA 98195, USA}

\date{\today}

\begin{abstract}
A complete mathematical theory of nonequilibrium thermodynamics of
stochastic systems in terms of master equations is presented.  As
generalizations of isothermal entropy and free energy, two functions
of state play central roles: the Gibbs entropy $S$ and the relative
entropy $F$, which are related via the stationary distribution of
the stochastic dynamics. $S$ satisfies the fundamental entropy
balance equation $dS/dt=e_p-h_d/T$ with entropy production rate
$e_p\ge 0$ and heat dissipation rate $h_d$, while $dF/dt=-f_d\le 0$.
For closed systems that satisfy detailed balance: $Te_p(t)=f_d(t)$.
For open systems one has $Te_p(t)=f_d(t)+Q_{hk}(t)$ where the
housekeeping heat, $Q_{hk}\ge 0$, was first introduced in the
phenomenological nonequilibrium steady state thermodynamics.
Entropy production $e_p$ consists of free energy dissipation
associated with spontaneous relaxation (i.e., self-organization), 
$f_d$, and active energy pumping that sustains the open system 
$Q_{hk}$. The amount of excess heat involved in the relaxation 
$Q_{ex}=h_d-Q_{hk} = f_d-T(dS/dt)$.

\end{abstract}

\maketitle

    Thermodynamics is the mathematical theory that describes
the transformations of energy among all its forms. In particular,
entropy and free energy are key quantities in understanding energy
transformations at finite temperature.  J.W. Gibbs' statistical
mechanics, which established the relation between molecular systems
and their equilibrium thermodynamics, is founded on the assumption
that isothermal molecular systems are stochastic, and macroscopic
quantities such as energy and number of particles can, and should be
treated as random variables with distributions in terms of the
concept of ensemble.  In Gibbs' approach, the fundamental origin of
the stochasticity, in the surrounding ``bath'', is left for others
to ponder, while he spearheaded into applications of his approach to
complex physical systems such as chemical solutions, systems too
complex to be rigorously studied by the kinetic theory of Boltzmann
\cite{khinchin}.

    The Gibbs' approach can be naturally generalized in two different
directions: The relaxation dynamics of a system in contact with an
equilibrium bath (iso-thermal, iso-chemical potential, etc.) and the
steady state of a system in contact with a nonequilibrium
environment with sustained chemical potential difference (i.e., an
open chemical systems \cite{NP,QH06}). In recent years, it
becomes increasingly clear that both theories can be framed in terms
of the mathematical theory of Markov processes
\cite{Hi_books,Sch,JQQ}. The significant progresses in both
fluctuation theorems in terms of Markov models and 1-dimensional
exclusion processes in terms of interacting particle systems testify
the centrality of the probability theory \cite{evans_02,derrida_98}.

    A thermodynamics theory has emerged from
the mathematical analysis of Markov processes \cite{qqt02}.
The present paper gives the first complete account of this emergent
mathematical structure. In addition, we also raise several issues
concerning the nature of dissipation, thus, irreversibility, and
its connection to measurement of heat.

    To avoid intricate mathematical techniques, we consider Markov
systems with discrete state variables and continuous time,
characterized by master equations.  The 1976 review  \cite{Sch}
provides a natural starting point.  Generalization of the our present
results to continuous variables in terms of Fokker-Planck equations is
straightforward.

\section{Master equations with detailed balance}

    We shall consider a molecular system in terms of
a Markov model, $dp_i(t)/dt = \sum_{j} \left(p_j q_{ji}
-p_iq_{ij}\right)$, that is irreducible: Hence there is a unique
long-time stationary probability distribution $\{p^e_i\}$. One class
of master equations is particularly important: Its stationary
distribution satisfies detailed balance: $p^e_iq_{ij}=p^e_jq_{ji}$,
where $q_{ij}$ is the transition probability rate from state $i$ to
$j$.

    The following results are well known:

(i) The system has an internal energy $u_i = -T\ln p^e_i$
\cite{footnote1}, and one can define an entropy $S$, a total
internal energy $U$, and a free energy $F$, all functions of the
state of the system
\begin{eqnarray}
    S[\{p_i\}] &=& -\sum_i p_i\ln p_i, \ \
    U[\{p_i\}] = \sum_i p_i u_i, \nonumber\\
    F &=& U-TS = T\sum_i p_i\ln\left(
        \frac{p_i}{p^e_i}\right).
\end{eqnarray}
Throughout the present work, $k_B=1$ and the temperature $T$
is assumed to be a constant.

(ii) For $p_i(t)$ as the time-dependent solution to the master
equation, all $U$, $S$ and $F$ are functions of $t$, and
\begin{equation}
    \frac{dU[\{p_i(t)\}]}{dt} = -h_d
    = -\sum_{i>j} \left(p_iq_{ij}-p_jq_{ji}\right)
        (u_i-u_j).
\label{hdr}
\end{equation}
\begin{equation}
    \frac{1}{T}\frac{dF[\{p_i(t)\}]}{dt} = -e_p
    = -\sum_{i>j} \left(p_iq_{ij}-p_jq_{ji}\right)
        \ln\left(\frac{p_iq_{ij}}{p_jq_{ji}}\right).
\label{epr}
\end{equation}
\begin{equation}
    T\frac{dS[\{p_i(t)\}]}{dt} = Te_p(t) - h_d(t).
\label{dS_dt}
\end{equation}
(\ref{hdr})-(\ref{dS_dt}) have very clear thermodynamic meanings:
The total internal energy of the system changes due to exchange heat
with the bath \cite{footnote2}, and the free energy of an isothermal
system spontaneously decreases until it reaches to its minimum at
equilibrium.  It can be shown that $F[\{p_i\}]\ge 0$.  The term
$e_p$ is called entropy production rate; it is the same are {\em
free energy dissipation rate}. The entropy of the system increases
due to entropy generated in spontaneous processes and decreases when
heat is expelled into the surrounding.  Note in isothermal system,
heat and entropy are related by a simple factor (temperature) due to
the Clausius equality.

    Therefore, there is a complete satisfying time-dependent
thermodynamics for system in contact with an equilibrium bath
relaxing to equilibrium. ``In contact with an equilibrium bath'' is
mathematically represented by the Wegscheider-Kolmogorov cycle
conditions on the $q_{ij}$'s \cite{JQQ,QH06}.

\section{Two Generalizations of the Thermodynamics to
Systems without Detailed Balance}

    Many of the above results can be generalized to
master equations without detailed balance.  The only relation that
no longer exists is detailed balance: $p^s_i/p^s_j \neq
q_{ij}/q_{ij}$ where we denote the unique steady state probability
distribution as $\{p^s_i\}$. Again, we can still phenomenologically
introduce $u_i=-T\ln p_i^s$ \cite{footnote3}, total energy $U=\sum_i
p_iu_i$, Gibbs entropy $S=\sum_i p_i\ln p_i$, and free energy
$F=U-TS =T\sum_i p_i\ln\left(p_i/p_i^s\right)$. The last term is
widely known as relative entropy.

{\bf\em Thermodynamics based on Gibbs' entropy.}
The Gibbs' entropy is a generalization of Boltzmann's formula
to situations with nonuniform probability distributions.
According to the the fundamental postulate
of nonequilibrium thermodynamics of de Groot and Mazur \cite{Groot62},
the entropy change $dS$ can be
distinguished between two terms: $d_eS$
is the transfer of entropy across the boundaries of the system, and
$d_iS$ is the entropy produced within the system due to
spontaneous processes.  In terms of the $q_{ij}$, we thus have
for isothermal systems \cite{NP,QH06}
\begin{equation}\label{EntropyEq}
\frac{dS(\{p_i(t)\})}{dt}=\frac{d_iS}{dt}+\frac{d_eS}{dt}=e_p(t)
        -\frac{h_d(t)}{T},
\end{equation}
where $e_p(t)=\frac{1}{2}\sum_{i,j} \left(p_iq_{ij}-p_jq_{ji}\right)
        \ln\left(\frac{p_iq_{ij}}{p_jq_{ji}}\right)$ is  the
instantaneous entropy production rate, and $h_d(t)=\frac{T}{2}\sum_{i,j}
\left(p_iq_{ij}-p_jq_{ji}\right)\ln\left(\frac{q_{ij}}{q_{ji}}\right)$
is the heat dissipation rate.
The $e_p$ is the focus of recent intense studies on fluctuation
theorem \cite{evans_02}.

{\bf\em Thermodynamics based on relative entropy.}
It is widely known in the mathematical literature that $F\ge 0$ and
$dF(\{p_i(t)\})/dt \le 0$, both are hold for master equation
without detailed balance \cite{Sch}.  In fact, in terms
of the $q_{ij}$, the {\em free energy dissipation} $f_d$
\begin{eqnarray}
\frac{dF(\{p_i(t)\})}{dt} &=&-f_d
\label{F_as_L}\\
    &=& \frac{T}{2}\sum_{i,j}
    \left(p_i(t)q_{ij}-p_j(t)q_{ji}\right)
        \ln\left(\frac{p_j(t)p_i^s}{p_i(t)p_j^s}\right).
 \nonumber
\end{eqnarray}
Many people have noticed this nice Lyapunov properties:
\cite{Ber_Leb_55} used it in their $H$-theorem for
stochastic dynamics, \cite{Sch} discussed it
in connection to Prigogine-Glansdorff's
criterion of steady-state thermodynamic stability.
Both Reguera, Rubi and Vilar, and Ao have constructed 
their thermodynamic theories, the mesoscopic nonequilibrium 
thermodynamics and the Darwinian dynamics respectively,
based on Eq. \ref{F_as_L}.  Mackey has presented the 
dynamic origin of the arrow of time based on 
$F$ \cite{qqt02}.

{\bf\em Relationship between $e_p$ and $f_d$ and decomposition of
heat dissipation.} Without the detailed balance, $f_d$ and $e_p$
are no longer identical, even though both have many important
properties. Both are nonnegative; both follow their own
fluctuation theorems \cite{ECM,Ge_2010}

    We discover that the difference between $e_p$ and $f_d$ is
in fact the ``housekeeping heat'',
\begin{equation}
    Q_{hk} = Te_p - f_d = \frac{T}{2}\sum_{i,j}
    \left(p_iq_{ij}-p_jq_{ji}\right)
        \ln\left(\frac{p_i^sq_{ij}}{p_j^sq_{ji}}\right),
\label{Qhk}
\end{equation}
first put forward by Oono and Paniconi in a purely phenomenological theory
of nonequilibrium steady state (NESS) thermodynamics \cite{OP}.
 They
decomposed the total heat dissipation into a ``housekeeping'' part
and an ``excess'' part.  In their theory, the irreversibility of
a process converting work into excess heat when modulo
housekeeping heat.  Later, Hatano and Sasa combined this
concept with Langevin dynamics and established a deep connection
between the phenomenological NESS thermodynamics and the Jarzynski
equality \cite{OP}. One of us recently has generalized their
extended form of the Second Law to abstract Markov processes
\cite{Ge_PRE2009}.

    By convention, we shall take the sign of heat to be positive when it
flows from the system to its environment.  Then, our $h_d$ is the total
heat dissipation, and the excess heat
\begin{equation}
     Q_{ex} = h_d - Q_{hk} = h_d - Te_p + f_d
        = f_d-T\frac{dS[\{p_i(t)\}]}{dt},
\end{equation}
which represents the ``hidden'' heat term that is involved in the
``driving mechanism'' of the open system.  According to this
phenomenological view, a NESS by definition is not driven.
It is important to point out that the entire NESS thermodynamics
based on $Q_{hk}$ and $Q_{ex}$ is nonlocal: This can be seen
from the Eq. \ref{Qhk} which contains the infinitely long time
$\{p^s_i\}$ in its definition.  For stochastic system without
detailed balance, $\{p^s_i\}$ in principle can not be constructed
locally as done following the Boltzmann's Law.

We also have
\begin{equation}
    T\cdot e_p=f_d+Q_{hk} = -\frac{dF[\{p_i(t)\}]}{dt}
            + Q_{hk}.
\end{equation}
This implies that there are two different origins for the
total entropy production: $-\frac{dF(t)}{dt}$ is from
the spontaneous, non-stationarity and $Q_{hk}(t)$ from
the driven mechanism that sustains the nonequilibrium
environment.  They are both traditionally called
``nonequilibrium'' and were not distinguished before.
The decomposition of the two provides a deeper understanding
of irreversibility, as we shall show next.

\section{Different Mathematical Statements of The
Second Law Based on Nonnegative $e_p$ and $f_d$}

    The Second Law of Thermodynamics for isothermal system
is about decreasing free energy, not increasing entropy.
For open systems, we still have the two important
quantities satisfying
\begin{subequations}
\begin{eqnarray}
    T\frac{dS(t)}{dt} &=& Te_p(t) - h_d(t) = f_d(t) - Q_{ex}(t);
\\
    \frac{dF(t)}{dt} &=& Q_{hk}(t) - Te_p(t) = -f_d(t) .
\end{eqnarray}
\end{subequations}
This much is mathematically concrete.  Heuristically, we note that
$h_d=-TdS_e/dt$ is the heat dissipated via the boundaries to its
environment.  Hence, an experimental observer who resides ``inside''
the system might not be able to aware of the dissipative nature of
the ``driving mechanism''.  In that case, he/she might be able to
construct a ``conservative dynamics'' and a complete analogue to
equilibrium thermodynamics based on $f_d$. The $f_d$ explicitly
characterizes the spontaneous relaxation processes.  This is indeed
the thesis of Oono and Paniconi, Ao, and Wang \cite{qqt02,wang}.
However, to an observer outside the system, as most cell biologists
studying a cell, the relaxation is a form of self-organization
driven by the external energy.

 Even in the steady state, there are
still nonzero $e_p$ and $h_d$: There is a continuous useful energy
being pumped into the system, that sustains the system in a NESS. We
think the issue here is rather deep and philosophical; hence we
shall defer further discussions elsewhere.

    From the definition of $e_p$ given above,
it is clear that $e_p(t)\ge 0$ and the equality holds if and
only if the system has detailed balance.  Then, immediately
we have several inequalities:
\begin{subequations}
\label{Secondlaw}
\begin{eqnarray}
\frac{dS(t)}{dt}+\frac{h_d(t)}{T}&=& e_p(t)\geq 0,
\label{Secondlaw1}\\
\frac{dF(t)}{dt}-Q_{hk}(t)&=&-T\cdot e_p(t)\leq 0.\label{Secondlaw2}
\end{eqnarray}
\end{subequations}
Eq. \ref{Secondlaw1} is just the well-known Clausius inequality
($dS\geq -Q_{tot}/T$), which could be rectified through the
quasi-stationary process to obtain expressions for the entropy
produced ($dS$) as the result of heat exchanges ($Q_{tot}=h_d$). And
Eq. \ref{Secondlaw2} is a general version of the free energy
inequality for the amount of work performed on the system, since the
work values must then be consistent with the Kelvin-Planck statement
\cite{Finn93} and forbids the systematic conversion of 100\% heat to
work isothermally.

More precise, the quantity $Q_{hk}(t)$ in Eq. \ref{Secondlaw2}
vanishes when the detailed balance condition,
$p^{s}_i(t)q_{ij}(t)=p^{s}_j(t)q_{ji}(t)$, holds and then it returns
to the traditional Helmhotz or Gibbs free energy criterion of
equilibrium thermodynamics, depending on whether a it is a $NVT$ or
$NPT$ system \cite{BQ2008}.

According to the nonnegativity of $f_d$, one has \cite{Ge_PRE2009}
\begin{subequations}
\label{Secondlaw_ex}
\begin{eqnarray}
&& T\cdot\frac{dS(t)}{dt}+Q_{ex}(t)= f_d(t)\geq 0,
\label{Secondlaw1_ex}\\
&& \frac{dF(t)}{dt} =-T\cdot e_p(t)+Q_{hk}(t)\leq
    0.\label{Secondlaw2_ex}
\end{eqnarray}
\end{subequations}
Eq. \ref{Secondlaw1_ex} is the extended form of Clausius
inequality during any nonequilibrium time-dependent process, which
is a specail case of Hatano and Sasa's latest work \cite{OP}. And
Eq. \ref{Secondlaw2_ex} provides an alternative form of free energy
inequality.

The housekeeping heat is also nonnegative, which implies the driven
nature of the system and it is in fact a measure of how far the
system is kept away from detailed balance. This measure is defined
not only to NESS, but also for spontaneous relaxation to NESS.

\section{Discussion}

        Equilibrium statistical mechanics of Gibbs is one
of the corner stones of theoretical physics.  Classical
thermodynamics, however, is applicable to both equilibrium
and nonequilibrium systems. In fact the most celebrated
statement of The Second Law of Thermodynamics concerns
with an irreversible, macroscopic, spontaneous molecular
process.

        There are two types of nonequilibrium processes
which have been widely studied: those of time-dependent nature and
those with stationary characteristics.  The theories of transport
and of Zwanzig-Kubo belong to the first type.  The irreversibility
of this type of processes is originated in their initial conditions.
They are of transient nature.  The second type requires a sustained,
continuous energy supply; it is driven.  One of the most important
examples is a living cell in its homeostasis \cite{BQ2008}. This
type of systems have been called {\em nonequilibrium steady state}
(NESS) \cite{JQQ}. Oono and Paniconi have developed a rather
rigorous phenomenological NESS thermodynamics \cite{OP}. One of the
most important concepts from their analysis is the decomposition of
total dissipated heat into $Q_{hk}$ and $Q_{ex}$.

    Following the tradition of Gibbs, an investigation of
thermodynamics of molecular systems has to be based on the
mathematical theory of stochastic processes.  The recently developed
stochastic thermodynamics has extended the field much further than
other approaches \cite{MLN86,Seifert}. There are fundamental
differences in the irreversible thermodynamics of driven systems and
irreversibility of a spontaneous process.  An important concept, the
{\em housekeeping heat}, has emerged in recent studies of such
systems and processes.  For non-driven isothermal systems, also
widely known as closed system, the free energy dissipation, which
characterizes spontaneous irreversibility, and the entropy
production, which characterizes the total irreversibility, are the
same. However, for driven (open) system, there is a difference
between these two quantities: The housekeeping heat is a measure of
the difference.



\small
\begin{thebibliography}{99}
\bibitem{khinchin}
J.W. Gibbs, {\em The Scientific Papers of J. Willard Gibbs}
(Dover, New York, 1961);
A.I. Khinchin, {\em Mathematical Foundations of Statistical Mechanics}
(Dover, New York, 1960);
L. Boltzmann, {\em Lectures on Gas Theory}, Translated by S.G. Brush
(Univ. Calif. Press, Berkeley, 1964).

\bibitem{NP}
G. Nicolis and I. Prigogine, {\em Self-organization in Nonequilibrium
Systems: From Dissipative Structures to Order Through Fluctuations}
(Wiley, New York, 1977);
H. Qian, Annu. Rev. Phys. Chem. {\bf 58}, 113 (2007).

\bibitem{QH06}
H. Qian, J. Chem. Phys. {\bf 110}, 15063 (2006).
H. Qian, J. Phys. Cond. Matt. {\bf 17}, S3783 (2005).

\bibitem{Hi_books}
T.L. Hill, {\em Free Energy Transduction in Biology}
(Academic, New York, 1977);
{\em Free Energy Transduction and Biochemical Cycle Kinetics}
(Springer, New York, 1989).

\bibitem{Sch}
J. Schnakenberg, Rev. Mod. Phys. {\bf 48}, 571 (1976).

\bibitem{JQQ}
D.Q. Jiang, M. Qian, M. and M.P. Qian, {\em Mathematical Theory of
Nonequilibrium Steady States}, LNM 1833 (Springer-Verlag, Berlin, 2004);
P. Gaspard, J. Chem. Phys. {\bf 120}, 8898 (2004).

\bibitem{evans_02}
D.J. Evans and D.J. Searles, Adv. Phys.  {\bf 51}, 1529 (2002);
E.M. Sevick, R. Prabhakar, S.R. Williams and D.J. Searles,
Ann. Rev. Phys. Chem. {\bf 59}, 603 (2008).

\bibitem{derrida_98}
B. Derrida, Phys. Reports {\bf 301}, 65 (1998).

\bibitem{qqt02}
H. Qian, Meth. Enzymol. {\bf 467}, 111 (2009);
P. Ao, Comm. Theoret. Phys. {\bf 49}, 1073 (2008);
D. Reguera, J.M. Rubi and J.M.G. Vilar, J. Phys. Chem. B. 
{\bf 109}, 21502 (2005);
H. Qian, M. Qian and X. Tang, J. Stat. Phys. {\bf 107}, 1129 (2002);
M.C. Mackey, {\em Time's Arrow: The Origins of Thermodynamic
Behavior} (Springer, New York 1992).

\bibitem{footnote1}
Internal energy defined this way uses the equilibrium free energy of
the system as the reference point.  If one has the internal energy
$\widetilde{u}_i$ given {\em a priori} from mechanics, then $u_i =
\widetilde{u}_i+\ln Z$ where $Z=\sum_i e^{-\widetilde{u}_i}$ is
Gibbs' canonical partition function.

\bibitem{footnote2}
If a system, like a molecular motor, also does work against its
surrounding, then the $h_d$ is the energy dissipation rate (edr)
that contains both heat dissipation as well as the work done.


\bibitem{Groot62} de Groot, S. R. and Mazur, P.: {\em Non-equilibrium thermodynamics}
North-Holland 1962

\bibitem{footnote3}
More precisely, we are interested in a system that is in contact
with an isothermal bath with chemical potential difference.  This is
a generalization of Gibbs' grand canonical ensemble. The $T$, thus,
is taken from the temperature of the bath.

\bibitem{Ber_Leb_55}
P.G. Bergmann and J.L. Lebowitz, Phys. Rev. {\bf 99}, 578 (1955);
H. Qian, Phys. Rev. E. {\bf 63}, 042103 (2001).

\bibitem{ECM}
D.J. Evans, E.G.D. Cohen and G.P. Morriss, Phys. Rev. Lett. {\bf 71}, 2401 (1993);
J. Kurchan, J. Phys. A: Math. Gen. {\bf 31}, 3719 (1998);
J.L. Lebowitz and H. Spohn, J. Stat. Phys. {\bf 95}, 333 (1999);
H. Ge and D.Q. Jiang, J. Phys. A. {\bf 40} F713 (2007).

\bibitem{Ge_2010}
H. Ge, Extended form of the Seconld Law and its fluctuation theorem for
time-dependent multi-dimensional diffusion processes.  In preparation.

\bibitem{OP}
Y. Oono and M. Paniconi, Prog. Theor. Phys. Suppl. {\bf 130}, 29 (1998);
T. Hatano and S. Sasa, Phys. Rev. Lett. {\bf 86}, 3463 (2001).

\bibitem{Ge_PRE2009} H. Ge, Phys. Rev. E {\bf 80}, 021137 (2009)

\bibitem{wang}
J.-T. Wang, {\em Nonequilibrium Nondissipative Thermodynamics}
(Springer, New York, 2002).

\bibitem{Finn93}
C.B.P. Finn,  {\em Thermal Physics,} 2nd ed. (Chapman and Hall, London, 1993).

\bibitem{BQ2008}
D.A. Beard  and H. Qian,
{\em Chemical Biophysics: Quantitative Analysis of Cellular
Systems} (Cambridge University Press, 2008).

\bibitem{MLN86}
C.Y. Mou, J.L. Luo and G. Nicolis, J. Chem. Phys. {\bf 84}, 7011 (1986). 

\bibitem{Seifert}
U. Seifert, Phys. Rev. Lett. {\bf 95}, 040602 (2005); T. Schmiedl
and U. Seifert, J. Chem. Phys. {\bf 126}, 044101 (2007); U. Seifert,
Eur. Phys. J. B. {\bf 64}, 423 (2008).

\end{thebibliography}
\end{document}